# Superconductivity in Ru substituted polycrystalline BaFe$_{2-x}$Ru$_x$As$_2$

Shilpam Sharma, A. Bharathi[*], Sharat Chandra, Raghavendra Reddy[#], S. Paulraj[&], A. T. Satya, V. S. Sastry, Ajay Gupta[#] and C. S. Sundar

Condensed Matter Physics Division, Materials Science Group, IGCAR, Kalpakkam, India, 603102

[#]UGC-DAE CSR, Indore, India 452017,

[&] Physics Department, Periyar University, Salem, India 636 011

Abstract

The occurrence of bulk superconductivity at ~22 K is reported in polycrystalline samples of BaFe$_{2-x}$Ru$_x$As$_2$ for nominal Ru content in the range of x=0.75 to 1.125. A systematic suppression of the spin density wave transition temperature (T$_{SDW}$) precedes the appearance of superconductivity in the system. A phase diagram is proposed based on the measured T$_{SDW}$ and superconducting transition temperature (T$_C$) variations as a function of Ru composition. Band structure calculations, indicate introduction of electron carriers in the system upon Ru substitutiom. The calculated magnetic moment on Fe shows a minimum at x=1.0, suggesting that the suppression of the magnetic moment is associated with the emergence of superconductivity. Results of low temperature and high field Mossbauer measurements are presented. These indicate weakening of magnetic interaction with Ru substitution

*Corresponding author
A.Bharathi
Condensed Matter Physics Division,
Materials Science Givision
Indira Gandhi Centre for Atomic Research
Kalpakkam. 603102. India
Tel ++91 44 27480081 Fax: ++91 44 27480081
bharathi@igcar.gov.in

# 1. INTRODUCTION

Over the last two years much progress has been made in establishing superconductivity unambiguously in $MFe_2As_2$ (M=Ba,Sr,Ca,Eu) systems [1-6]. The pristine sample that has a Spin Density Wave (SDW) ground state is nudged into a superconducting (SC) state by electron/ hole doping and application of pressure [2,7,8,9,10]. Band structure calculations point to the fact that SDW state arises on account of the special 2D geometry of Fermi surface that is unstable to nesting [11,12]. Also associated with or preceding the magnetic transition is a tetragonal to orthorhombic structural transition, which is suppressed in the superconducting state. The strong interplay of structure, magnetism and electronic structure have been investigated recently in the Co substituted $BaFe_{2-x}Co_xAs_2$ system[13]. The temperature composition phase diagrams determined for the different chemical substitutions at different sites in $BaFe_2As_2$ [14,15,16] show a generic behaviour as a function of the concentration of the substituent, viz., a systematic suppression of the SDW transition, followed by co-existence of SDW and SC and the occurrence of a superconducting dome. Several transition metal (TM) substitutions with electrons in excess of Fe forming, $BaFe_{2-x}TM_xAs_2$ have been studied but the maximum $T_C$ has remained at ~25 K [17]. A much higher $T_C$ of 38 K and ~35 K, were however observed by optimal hole doping in the $Ba_{1-x}K_xFe_2As_2$ [2] system and in $BaFe_2As_2$ by application of high pressure[10]. A systematic investigation on the role of hydrostaticity, in the pressure dependent resistivity study of $BaFe_2As_2$, has revealed that uniaxial pressure favours the occurrence of high $T_C$ at 36 K whereas a lower $T_C$ of 29 K occurs under truly hydrostatic pressure [18]. Consistent with this finding are results that indicate that strained crystals of $BaFe_2As_2$ and $SrFe_2As_2$ display superconductivity at ambient pressure [19]. A compilation of structural data from several compounds of the related ReOFeAs (Re=rare-earth) superconducting family, indicates that $T_C$ is optimized at a particular Fe-As distance[20] and/or at a particular Fe-As tetrahedral angle[21], indicating that the local structure of the $FeAs_4$ tetrahedra plays a crucial role in determining $T_C$. Devising schemes to effect structural distortions by chemical substitution that would lead to higher $T_C$ in the $BaFe_2As_2$ system will be useful.

Thus motivated, we examine the effect of Ru substitution at the Fe site in $BaFe_2As_2$. At the outset, it is clear that Ru is isoelectronic to Fe and being larger in size should introduce steric effects, affecting the Fe-As bond length leading to distortions of the FeAs$_4$ tetrahedral motifs. In addition, the larger radius of the 4d electron shell should increase the metal-metal overlap in the Fe/Ru layer and increase the hybridization of metal atom with As leading to significant alterations in the electronic structure. It is established that BaRu$_2$As$_2$ forms by solid state reaction, is iso-structural to BaFe$_2$As$_2$ and is metallic although non-superconducting[22], indicating the feasibility of Ru substitution in BaFe$_2$As$_2$. Here we report on the synthesis of polycrystalline samples of the BaFe$_{2-x}$Ru$_x$As$_2$ series for various Ru fraction, x, and on investigations of their structural, magnetic and superconducting properties. The study indicates a systematic suppression of the low temperature SDW state with increase in Ru concentration, leading to the observation of superconductivity in BaFe$_{2-x}$Ru$_x$As$_2$ at x~1.0, at ~20 K. Notably, at this composition the magnetic moment at Fe is suppressed as verified by Mossbauer measurements and substantiated by band structure calculations.

## 2. EXPERIMENTAL DETAILS

The BaFe$_{2-x}$Ru$_x$As$_2$ (x=0.0, 0.25, 0.50, 0.75, 0.875, 1.0, 1.125, 1.25 and 1.5) samples were prepared by solid state reaction from preformed FeAs and RuAs powders and Ba chunks, under 30 bar Ar pressure [23]. After an initial heat treatment at 1233 K for 10 hours the reacted powder was ground, pelletised and sintered at 1173 K for 5 hours. All weighing operations and loading of the reactants into the Ta crucibles were done in a Helium filled glove box. The FeAs and RuAs powders were prepared in the same set up by heat treating the intimate mixtures of Fe and As powders in quartz crucibles in the temperature range of 873 K to 1073 K for 6 hours. The procedure was repeated twice with an intermediate grinding. The samples were characterized for phase formation and crystal structure using a STOE diffractometer operating in the Bragg-Brentano geometry. The resistivity measurements carried out in the four probe geometry, were done in a dipper cryostat. The diamagnetism of the samples was confirmed by magnetisation measurements in a CRYOGENIC, UK make liquid helium based vibrating sample magnetometer operating at 20.4 Hz. H$_{C2}$ and the Hall co-efficient measurements were carried out in an exchange gas cryostat in the 6 K to 300K temperature range under magnetic fields upto 12 Tesla. $^{57}$Fe Mossbauer measurements were carried out in transmission mode with $^{57}$Co radioactive source in constant acceleration mode using standard PC-based Mossbauer spectrometer equipped with a Weissel make velocity drive. The measurements were carried out at 300 K, 5 K and 5 Tesla external magnetic field applied parallel to the gamma rays (using JANIS SuperOptiMag superconducting magnet). Velocity calibration of the spectrometer was done with natural iron absorber at room temperature.

## 3. RESULTS & DISCUSSIONS

The x-ray diffraction results indicate the formation of the I4/mmm, ThCr$_2$Si$_2$ structure for all Ru fractions substituted. A small fraction of impurity peaks due to Ru/Fe and Ba arsenides were identified. From Rietveld analysis of the XRD data obtained for the x=0.875 sample, the fractional z-coordinate of Arsenic ($Z_{As}$), transition metal-As bond length, the two ($e_{1,2}$) tetrahedral angles of FeAs$_4$ tetrahedron were determined to be, 0.3554, 0.2424 nm and 112.51 degrees and 107.97 degrees respectively. The refined Ru composition was determined to be 0.864, which is close to the nominal composition. The tetrahedral angles are similar to those obtained in the case of iso-electronic substitution of P at As site for the optimal superconducting composition [16], but different from those seen under the application of pressure[24].

The lattice parameter variations obtained from XRD data as a function Ru concentration are shown in Fig.1. It is clear from the figure that with Ru substitution there is an increase in the a-lattice parameter, whereas the c-lattice parameter shows a decrease, leading to a decrease in the c/a ratio. There is an overall increase in the cell volume of ~1.18% for x=1.0. Similar changes were observed in the SrFe$_{2-x}$Ru$_x$As$_2$ system[25]. The changes in the cell parameters due to Ru substitution contrasts with the variations obtained in Co [26] and P substituted BaFe$_2$As$_2$ [16] and that under the application of pressure[24], where a monotonic decrease in cell volume arises as a consequence of a decrease in both the a and c lattice parameters. In the Ba$_{1-x}$K$_x$Fe$_2$As$_2$ system, while the a- lattice parameter decreases and c-lattice parameter increases, the cell volume remains constant with substitution [27].

The temperature dependence of resistivity normalized to the room temperature value obtained in all Ru substituted samples are shown in various panels in Fig.2. The well known drop in resistivity corresponding to the SDW transition seen in BaFe$_2$As$_2$ [1] is clearly visible at ~ 150 K in the x=0.0 sample. With increasing Ru substitution viz., for x=0.25 0.50 and 0.625, the room temperature resistivity decreases and the SDW transition shifts to a lower temperature. The onset of the SDW transition for the different Ru compositions are marked by '*' in the figure. For a Ru fraction of x=0.75, a small bump due to the SDW transition is seen. In addition the resistivity in this sample shows a clear signature of the occurrence of a superconducting transition with an onset of 22 K, leading to zero resistance. This transition to the superconducting state is clearly seen in samples with Ru fraction of x=0.875, x=1.0 and x=1.125, but the anomaly due to SDW transition is not observed in them. The normal state resistance in these superconducting samples indicate a linear T dependence up to 250 K. Further for the x=0.625 and x=1.25 samples, although a fall in R(T)/R(300 K) is observed, no zero resistance is seen and in the sample with x=1.5 no drop in resistivity is observed. It can be seen from the figure that the normal state resistivity acquires curvature for the x=1.25 and 1.5 samples. For these samples the R(T) in the normal state was seen to fit to a $T^n$ power law with n~1.5. Similar composition dependent changes in the power law behaviour of the normal state R(T) was seen in the BaFe$_2$As$_{2-x}$P$_x$ system[16] and under the application of pressure[10]. The zero field cooled (ZFC) magnetization data of the four samples with Ru fraction x=0.75 to 1.125 sample is shown in Fig.3. The diamagnetic drop at ~20 K is evident from the data for samples with x in the range of x=0.75 to x=1.125. No diamagnetic signals were observed for samples with x = 0.625 and x =1.25. The presence of zero resistance (cf. Fig.2) and diamagnetism (cf. Fig.3) in the samples with Ru composition in the composition range of x=0.75 to x=1.125 provides unambiguous evidence for observance of superconductivity in the BaFe$_{2-x}$Ru$_x$As$_2$ system in this composition regime.

The variation of resistivity with temperature for the sample with a Ru fraction of x=0.875, measured under various fields upto 12 Tesla is shown in Fig.4. A systematic decrease in the superconducting onset with increasing magnetic field is clearly seen.

Broadening of the superconducting transitions upon the application of magnetic field is negligible indicating the minor role of anisotropy and granularity in this system. This behaviour is similar to results obtained in superconducting, $Ba_{1-x}K_xFe_2As_2$ [23] and $BaFe_{2-x}Co_xAs_2$[26]. A linear fit to the onset data resulted in the evaluation of $-dH_{C2}/dT$ at $T_C$ to be 2 T/K. This value is close to that obtained for Co substitution in $Ba(Fe_{1-x}Co_x)_2As_2$ [26] for x=0.034 and x=0.07, but is substantially smaller than the value of ~7 T/K seen [23] in K substituted samples having a $T_C$ of 38 K. The variation of the critical field versus normalized $T_C$ ($T_C(B)/T_C(0)$) are compared for the three superconducting samples with varying Ru compositions in the inset of Fig.4.

In Fig.5, we summarise the variation of $T_{SDW}$ and $T_C$ obtained from the resistivity curves shown Fig.2, as a function of Ru concentration. Data for samples that exhibit zero resistivity are only included in this figure. The superconducting onsets obtained from magnetization (cf. Fig.3) are also shown in the figure. It is evident from the figure that the SDW transition temperature decreases from that in the pristine compound for Ru fractions upto x=0.75. Co-existence of SDW and SC seems to occur in the x=0.75 sample. It is clear from Fig.5 that for Ru fractions of x=0.875, 1.0 and 1.125, only the superconducting phase is stabilized. Superconductivity is absent for Ru concentration greater than or equal to x=1.25.

To investigate how Ru substitution affects the electronic structure, we have performed spin polarized density functional calculations for $BaFe_2As_2$, $BaFe_{1.5}Ru_{0.5}As_2$, $BaFe_{1.0}Ru_{1.0}As_2$, $BaFe_{0.50}Ru_{1.5}As_2$, and $BaRu_2As_2$, using the full potential linearized plane wave plus localized orbitals (FP-LAPW+LO) method, with the WIEN2k code [27] and the results are shown in Fig.6. The details of the calculation is are elaborated in [28]. The calculations were carried out using a superstructure obtained by the $\sqrt{2}\times\sqrt{2}\times1$ construction (**a'** = (**a+b**), **b'** = (**a-b**), **c'** = **c**) from the crystal structure of $BaFe_2As_2$ [30], as shown in Fig.6c. In this reconstructed structure, all the four Fe atoms occupy non equivalent sites. The experimental lattice parameters of the Ru fraction of x = 0.5, x = 1.0 and x = 2.0 were used for the Ru substituted electronic structure calculations. The calculated DOS shown in Fig.6a for $BaFe_2As_2$ is in agreement with that obtained earlier [29]. The atom resolved DOS for *d*-Fe and *p*-As are also indicated in the figure. The Ba atom does not contribute appreciably to the DOS near the Fermi level ($E_F$). A steady increase in DOS at $E_F$ was seen with increase Ru substitution. This is clear from Fig.6b, where with 50% Ru substitution, the DOS at $E_F$ increases to 4.40 eV/unit cell/atom from 1.82 eV/unitcell/atom in the $BaFe_2As_2$ (cf Fig.6a). A similar increase in DOS at $E_F$ has been observed for the $BaFe_2As_2$ system under the application of pressure and with K doping [24]. The significant broadening and their increased contribution to DOS, suggests that the Fe 3*d* electrons get delocalized with Ru substitution. Ru *d* levels also contribute to a small extent to the DOS at $E_F$. The converged $E_F$ for $BaFe_2As_2$ is 0.60268 Ryd, while that for $BaFe_{1.5}Ru_{0.5}As_2$ is 0.63645 Ryd. The upward shift in $E_F$ with Ru addition implies an electron doping due to Ru substitution.

To check on the nature of carriers introduced due to Ru substitution, Hall co-efficient measurements were carried out in the 10 K to 300K temperature range in a home built set up in the Van der Pauw geometry[31]. The $R_H$ versus T for the single crystalline sample of $BaFe_2As_2$ and the $R_H$ versus temperature on the polycrystalline $BaFe_{2-x}Ru_xAs_2$ for the nominal x=0.75 sample are displayed Fig.7a and 7b respectively. Both the magnitude and temperature dependence of the Hall co-efficient shown in Fig.7a are in agreement with earlier reports on single crystals of $BaFe_2As_2$[32]. From Fig.7b, it is clear that the $R_H$ is negative in the normal state of the superconducting, Ru substituted sample. The $R_H$ values also shows the characteristic drop to zero due to occurrence of superconductivity below ~20 K. The $R_H$ value in the normal state of the Ru substituted sample shown in Fig.7b is smaller than that seen in $Ba_{0.65}K_{0.45}Fe_2As_2$ [31] and that in the $BaFe_{1.8}Co_{0.2}As_2$ system[8] and is temperature independent. Assuming a one band model, this small $R_H$ translates to an electron density, which is ~10 times as large as that evaluated for the $BaFe_{1.8}Co_{0.2}As_2$ sample[8].

Band structure calculations were also used to evaluate the evolution of the magnetic state of $BaFe_2As_2$ as a function of Ru substitution. For the x=0.0 and x=0.5 structures the lowest energy occurs for the stripe anti-ferromagnetic order[30] and for the x=1.0 and 1.5 superstructures the lowest energy configuration turns out to be paramagnetic. The magnetic moments of the Fe/Ru atoms were obtained from unconstrained minimization of total energy during the self consistent iterations in the spin polarized calculations. It was seen that the Ru atoms always align antiferromagnetically to the Fe atoms and only in the presence of the Fe atoms, do they show a small magnetic moment. In the x = 0.5 and 1.5 compositions, there are three Fe(Ru) atoms for one Ru(Fe), whereas, for x = 1.0, there are two Fe atoms for two Ru. Hence for the x = 0.5 and 1.5 compositions, the magnetic moments obtained for all the Fe atoms in the unit cell are different. The variation of the magnitude of Fe and Ru moment with increase in Ru concentration is shown in Fig.8, where the maximum of the moments obtained for the Fe/Ru atom are plotted. The Fe magnetic moment shows a minimum value at x = 1.0. A small contribution from the Ru magnetic moment is also evident from Fig.8. The Ru atoms in $BaRu_2As_2$ show zero magnetic moments. Fig.9 displays the evolution of the Fermi surface with Ru substitution, as obtained from band structure calculations for the spin-down bands. The spin-up bands also show similar features in all the cases excepting for x = 0.5. A few points to mention on careful perusal of Fig.9 are: (i) the Fermi surfaces become more connected with increase in Ru content, suggestive of increased delocalization of the carriers at $E_F$. (ii) Fermi surfaces for spin up and spin down electrons become very similar with increase in Ru content, suggesting a preference for a paramagnetic ground state for a substantial increase in Ru content and (iii) The Fermi surface gets larger, indicating electron addition due to Ru substitution.

To experimentally investigate the magnetic moment at the Fe site, Mossbauer measurements were carried out on the x=0.0, 0.5 and 1.0 samples at 5 K. The Mossbauer data for these samples obtained at 5 K are displayed in Fig.10. The Mossbauer data displayed for the x=0.0 and 0.5 samples (cf. Fig.10a,b), show the characteristic six finger pattern due to magnetic ordering in the spin density wave state of the samples at 5 K. The magnetic hyperfine split data is analyzed with NORMOS-DIST program[33] and the data of superconducting sample is analyzed with NORMOS-SITE program[33]. The hyperfine magnetic split spectrum is fitted to a distribution of fields for x=0 and 0.5 samples as shown by solid lines in the figure. The probability distribution of hyperfine fields obtained from the fits, for the x=0.0 and x=0.5 samples are shown in the inset of

Fig.10a. It is clear from the figure, that the hyperfine field distribution has a maximum at a field of ~ 5 T in the pristine sample and it becomes more spread out in the Ru containing sample with x=0.5. The observed value of average hyperfine field (BHF) for x=0 sample matches closely with that obtained earlier [34,35]. The Mossbauer spectrum of x=1.0 sample shown in Fig.10c shows a broad singlet which is similar to that reported in K substituted samples which show superconductivity at 38K [34]. The fitted data of Fig.10c has an isomer shift value of 0.55 ± 0.01 mm/sec corresponding to superconducting phase and a doublet with the hyperfine parameters matching with the $FeAs_2$ impurity phase [36]. The fraction of the $FeAs_2$ phase is estimated to be about ~3.5%. Fig. 10(d) shows the Mossbauer spectrum of x=1.0 sample measured at 5 K and 5 Tesla external magnetic field. The fact that the observed effective BHF (internal field) value is close to that of applied external magnetic field, within experimental errors, indicates that there is no magnetic ordering present in the x=1.0 sample. The Mossbauer results thus indicate the magnetic interactions due to the SDW phase weaken with Ru substitution and are altogether absent in the superconducting sample with Ru composition of x=1.0.

It will be instructive to compare the phase diagrams determined from our study for $BaFe_{2-x}Ru_xAs_2$ with those of other systems belonging to the $BaFe_2As_2$ class. Co-existence of SDW and SC phases, occur in all the phase diagrams irrespective of how the phase changes are induced, i.e: by pressure or substitution, isoelectronic or otherwise. Our study indicates that optimal $T_C$ occurs for Ru concentration x between 0.75 and 0.875. This is similar to the isoelectronic substitution cases, of $BaFe_2As_{2-x}P_x$ and $SrFe_{2-x}Ru_xAs_2$, in which that optimal superconductivity occurs for x ~0.8. Whereas, it contrasts with optimal $T_C$ being observed at x~0.2 in the more extensively studied, $BaFe_{2-x}TM_xAs_2$ [17] systems in which the TM atoms add extra electrons as compared to Fe. This difference emphasises the fact that to effect the electronic structure change conducive for superconductivity a much larger distortion of the lattice is required for the case of iso-electronic substitution.

## 5. SUMMARY & CONCLUSIONS

Polycrystalline samples of $BaFe_{2-x}Ru_xAs_2$ samples have been synthesized by solid state reaction from the FeAs, RuAs powders and Ba chunks under 30 bar argon pressure. Resistivity, DC magnetization, Hall co-efficient and Mossbauer measurements were employed to characterize the physical properties of the series. A phase diagram is proposed that indicates that the spin density wave ground state gives way to the occurrence of superconductivity in the x range of 0.75 to 1.125. Hall co-efficient measurements indicate introduction of electron carriers due to Ru substitution. Mossbauer results indicate a systematic suppression of the low temperature magnetic state with increase in Ru content. Electron doping and magnetic moment suppression with Ru substitution are borne out by band structure calculations.

## ACKNOWLEDGEMENT

The authors are deeply indebted to Dr. Y. Hariharan, a former member of the group for elaborate discussions and critical reading of the manuscript. S. Paulraj acknowledges financial support from UGC-DAE-CSR for the research fellowship and Dr. C Sekar of Periyar University Salem, for his encouragement during this work.

**References**

1. M. Rotter, M. Tegel, and D. Johrendt, Phys. Rev. B **78**, 020503 (2008)
2. M. Rotter, M. Tegel and D. Johrendt, Phys. Rev. Lett., **101**, 107006 (2008)
3. K. Sasmal, B.Lv, B.Lorenz, A.M.Guloy, F. Chen, Y. Y. Xue, C.W. Chu, Phys. Rev. Lett. **101**, 107007 (2008)
4. Z. Ren, Z. Zhu, S. Jiang, X. Xu, Q. Tao, C. Wang, C. Feng, G. Cao and Z.Xu, Phys. Rev. B **78**, 052501 (2008)
5. H.S. Jeevan, Z. Hossain, D. Kasinathan, H. Rosner, C.Geibel and P. Gegenwart, Phys. Rev. B 78, 092406 (2008)
6. M.S.Torikachvilli, S. L. Budko, N. Ni and P. C. Canfield Phys. Rev. Lett. **101**, 057006 (2008)

7. N. Ni, S. L. Bud'ko, A. Kreyssig, S. Nandi, G. E. Rustan, A. I. Goldman, S. Gupta, J. D. Corbett, A. Kracher, and P. C. Canfield, Phys. Rev. B, **78**, 014507 (2008)
8. A. S. Sefat, R. Jin, M. A. McGuire, B. C. Sales, D. J. Singh, and D. Mandrus, Phys. Rev. Lett.,**101,** 117004 (2008)
9. Patricia L Alireza, Y T Chris Ko, Jack Gillett, Chiara M Petrone, Jacqueline M Cole, Gilbert G Lonzarich and Suchitra E Sebastian, J. Phys: Condens. Matt. **21**, 012208 (2008)
10. A.Mani, N. Ghosh, S. Paulraj, A. Bharathi, C. S. Sundar, Euro. Phys. Lett. **87**, 17004 (2009)
11. D.J.Singh Physica C **469**, 418 (2009)
12. D. Kasinathan, A. Ormeci, K. Koch, U. Burkhardt, W. Schnelle, A. Leithe-Jasper, H. Rosner, New J. Phys. **11** 025023 (2009)
13. S. Nandi, M. G. Kim, A. Kressig, R. M. Fernades, D. K. Pratt, A. Thaler, N. Ni and S.L. Bud'ko, P. C. Canfield, J. Schamalian, R. J. McQueeney and A. I. Goldman, arXiv:cond-mat 0911.3136
14. D. Johrendt and R. Poettgen Physica C **469**, 332 (2009)
15. J.-H. Chu, J.G.Analytis, C. Kuchaerczy, I. R. Fisher, Phys. Rev. B **79**, 014506 (2008)
16. S.Jiang, H. Xing, G. Xuan, C. Wang, Z. Ren, C. Feng, J. Dai, Z. Xu and G. Cao, J.Phys. Condens. Matter **21,** 382203 (2009)
17. P. C. Canfield and S.L.Budko, arXiv:cond-mat 1002.0858
18. W.J. Duncan, O. P.Welzel, C. Harrison, X. F. Wang, X. H. Chen, F. M. Grosche and P. G. Niklowitz, arXiv:cond-mat 0910.4267
19. J. S. Kim, T. D. Blasius, E. G. Kim and G. R. Stewart, J. Phys.: Condens. Matter **21,** 342201 (2009)
20. Y. Mizuguchi, Y. Hara, K. Deguchi, S. Tsuda, T. Yamaguchi, K. Takeda, H. Kotegawa, H. Tou and Y. Takano arXiv:cond-mat 1001.1801
21. K. Ishida, Y. Nakai and H. Hosono, J. Phys. Soc. of Japan, **78**, 1 (2009)
22. R. Nath, Yogesh Singh, and D. C. Johnston, Phys. Rev. B **79**, 174513 (2009)

23. A.Bharathi, Shilpam Sharma, S. Paulraj, A. T. Satya, Y. Hariharan, C.S. Sundar, Physica C: **470**, 8 (2010)

24. S. A. J. Kimber, A. Kressig, Y.Z. Zhang, H. O. Jeschke, R. Valenti, F. Yokaichiya, E. Colombier, J. Yan, T. C. Hansen, T. Chatterji, R. J. McQueeney, P. C. Canfield, A. I. Goldman and D. N. Argyriou, Nature Materials, **8**, 471 (2009)

25. W. Schnelle, A. Leithe-Jasper, R. Gumeniuk, U. Burkhardt, D. Kasinathan, H. Rosner, Phys. Rev. B **79**, 214516 (2009)

26. N. Ni, M. E. Tillman, J.-Q. Yan, A. Kracher, S. T. Hannahs, S. L. Bud'ko, and P. C. Canfield, Phys. Rev. B **78** 214515 (2008)

27. P. Blaha, K. Schwarz, G. K. H. Madsen, D. Kvasnicka and J. Luitz, WIEN2k, An Augmented Plane Wave + Local Orbitals Program for Calculating Crystal Properties (Karlheinz Schwarz, Techn. Universitat Wien, Austria), 2001. ISBN 3-9501031-1-2

28. The generalized gradient approximation (PBE96) was used for the exchange interaction. The calculations were done with 1000 k-points in the full Brillouin zone and $R_{MT}*K_{max}$ was 12 for all the calculations. The ground state relaxed structure was obtained using the same method reported previously in literature [29] for the same structure. The muffin tin radii used were $2.2a_0$ for Ba and $2.1a_0$ for Fe, Ru and As, where $a_0$ is the Bohr radius [29]. The calculation was converged with respect to the energy, charge displacement and forces to the tune of $10^{-6}$ Rydberg, $10^{-6}$ Bohr and 1 mRydberg/Bohr.

29. D. J. Singh, Phys. Rev. B **78**, 94511 (2008)

30. Q. Huang, Y.Qiu, W. Bao, J. Lynn, M. Green, Y.Chen, T.Wu, G. Wu and X.Chen, Phys. Rev Lett. **101** 257003 (2008)

31. Shilpam Sharma, A. Bharathi, Y. Hariharan, C.S. Sundar, Proceedings of DAE solid State Physics Symposium India **53**, 917 (2008)

32. C. L. Zentile, J. Gillett, S. E. Sebastian, and J. R. Cooper;cond-mat

    arXiv:0911.1259v1

33. NORMOS: R.A.Brand., Universitaet Duisburg, 1990

34. I.Novik and I Felner Physica C **469** 485 (2009)

35. M.Rotter, M. Tagel, D. Johrendt, I. Schellenberg, W. Hermes and R. Poettgen Phys. Rev. B **78,** 020503 (R) (2008)].

36. Israel Felner, Israel Nowik, B.L.Joshua H.Tapp, Z.Tang and A.M.Guloy., Hyperfine Interactions **191,** 61 (2009)

**FIGURE CAPTIONS**

**Fig.1a** Variation of the a and c lattice parameter as a function of Ru fraction substituted.

**Fig.2** Variation in normalised resistivity with temperature in $BaFe_{2-x}Ru_xAs_2$ for various nominal Ru fractions x indicated. The origin is shifted for each composition for clarity. The ordinate axis alternates for each composition showing the 0 and 1 markers. The SDW transitions are indicated by stars.

**Fig.3** DC magnetisation as a function of temperature for nominal x=0.75, x=0.875, x=1.0 and x=1.125.

**Fig.4** The variation of the SDW transition temperatures and superconducting onsets as a function nominal Ru fraction x. SDW and superconductivity states co-exist for the x=0.75 sample.

**Fig.5** Variation of resistance versus temperature for various external magnetic fields indicated, for Ru fraction, x= 0.875. Inset shows the field dependence of superconducting transition temperatures, normalized to their zero field counterparts, for x=0.75,0.875 and 1.125.

**Fig.6** A comparison of the spin polarized density of states obtained from first principles calculations in (a)$BaFe_2As_2$ and (b) $BaFe_{1.0}Ru_{1.0}As_2$. (c) a schematic of the structure of the supercell used in the calculation

**Fig.7a** Hall co-efficient in $BaFe_2As_2$ single crystals (b) Hall co-efficient in polycrystalline $BaFe_{2-x}Ru_xAs_2$ for x=0.75; the measuring field and current used in the experiments are indicated.

**Fig.8a** The variation of the calculated magnetic moment at the Fe site and Ru site with increase in Ru concentration in the supercell calculations.

**Fig.9** Evolution of the Fermi surface obtained from spin polarized DFT calculations for the spin up bands in $BaFe_{2-x}Ru_xAs_2$ for (a) x=0.0, (b) x=0.5, (c) x=1.0 and (d) x=2.0.

**Fig.10** Variation of the transmitted intensity versus the velocity, of the Mossbuer spectra measured at 5 K in $BaFe_{2-x}Ru_xAs_2$ samples with (a)x=0.0, (b)x=0.5 (c)x=1.0 and (d) x=1.0 with an external magnetic field of 5 T. Inset in (a) shows the probability distribution of the magnetic field for x=0.0 and x=0.5 samples, obtained from fits of the corresponding data shown in (a) and (b).

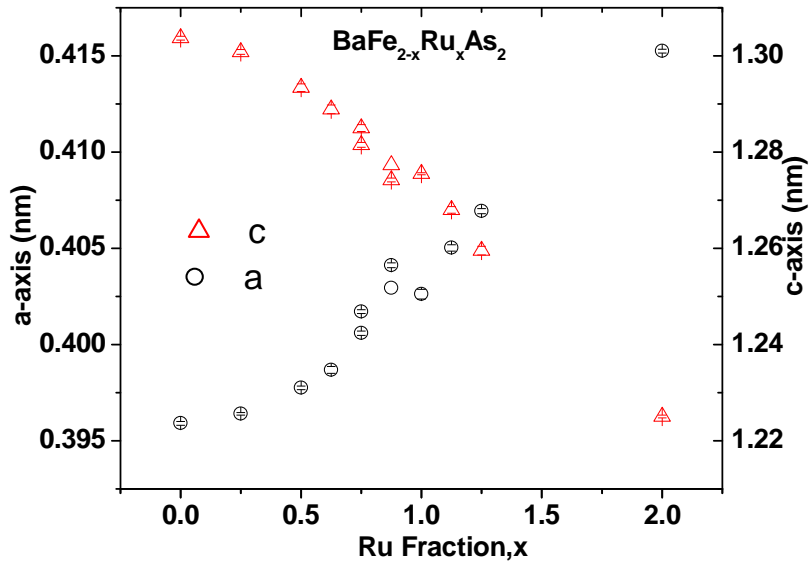

Fig.1

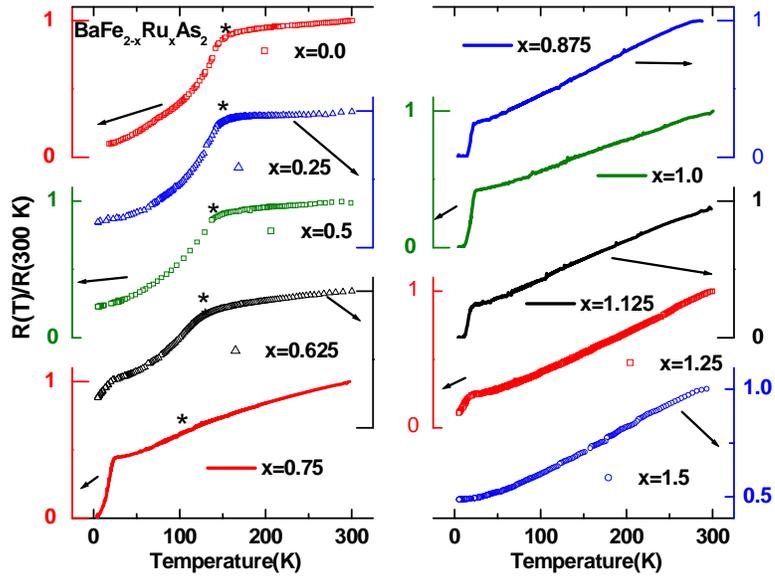

Fig.2

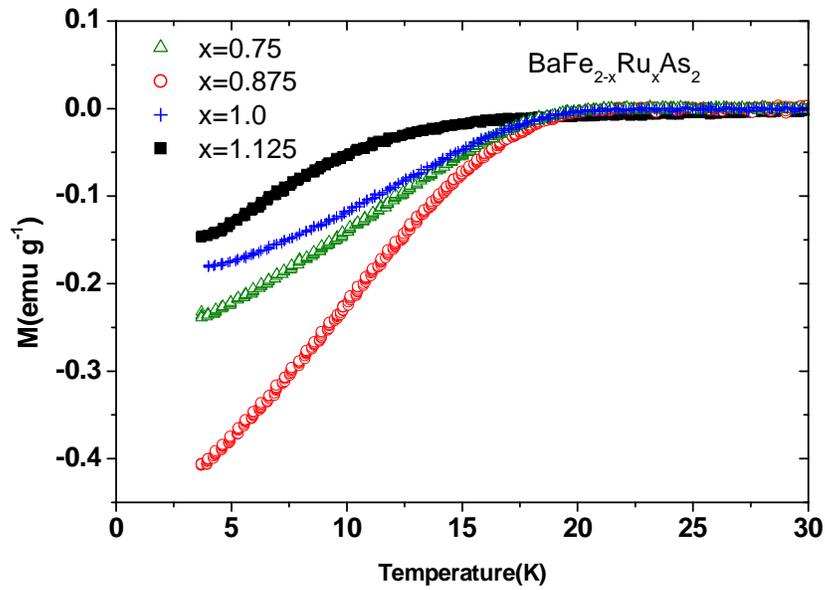

Fig.3

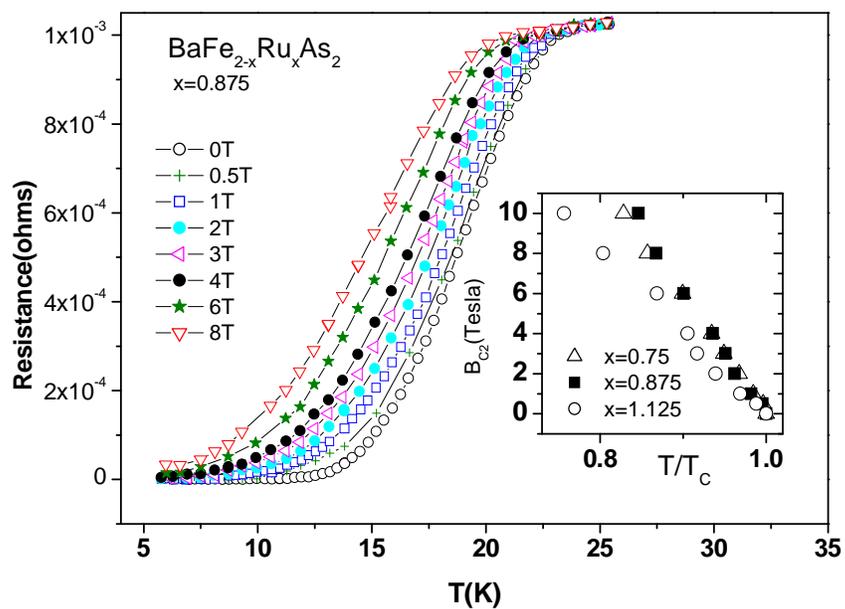

Fig.4

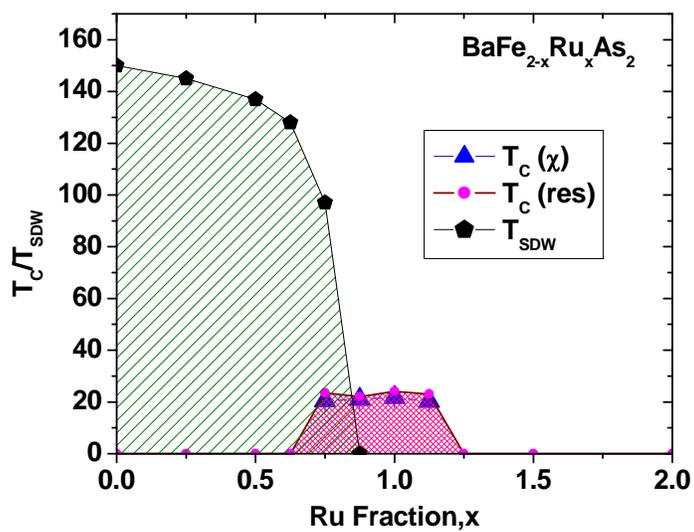

Fig.5

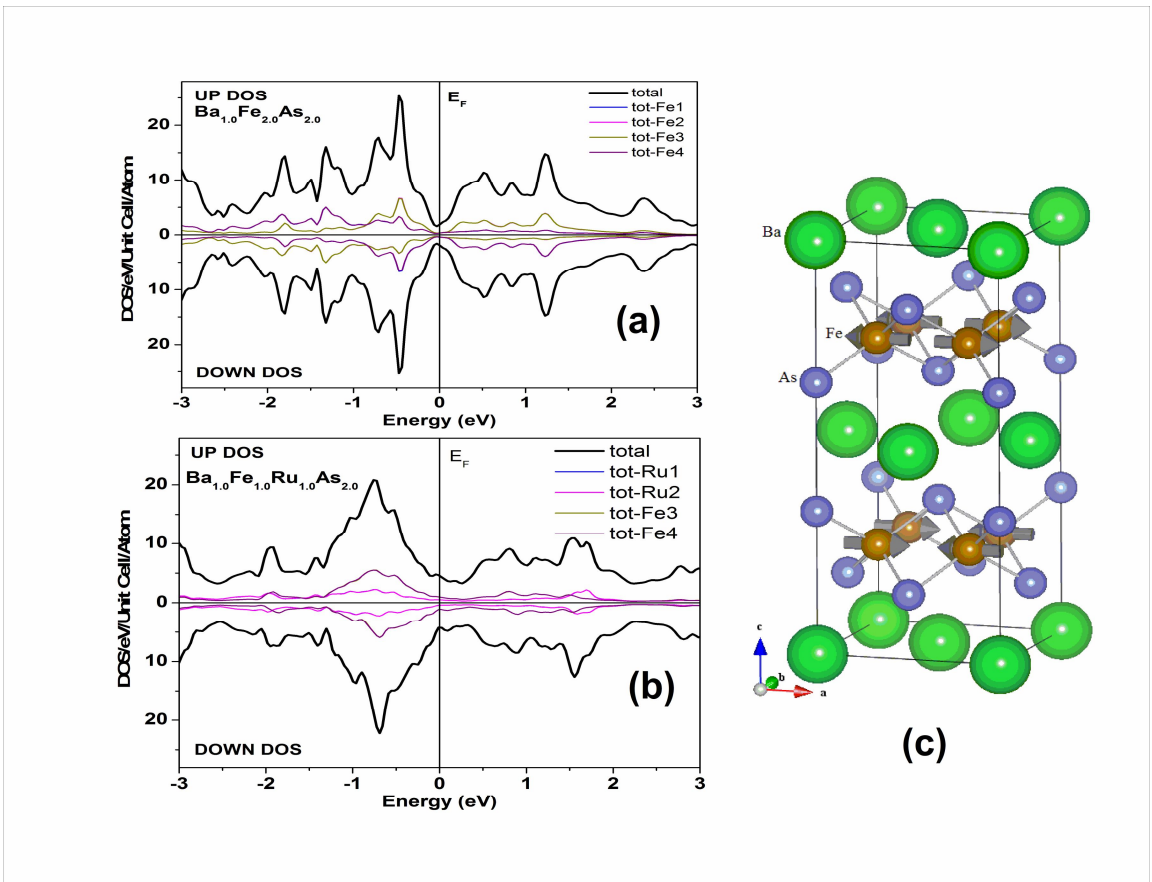

Fig.6

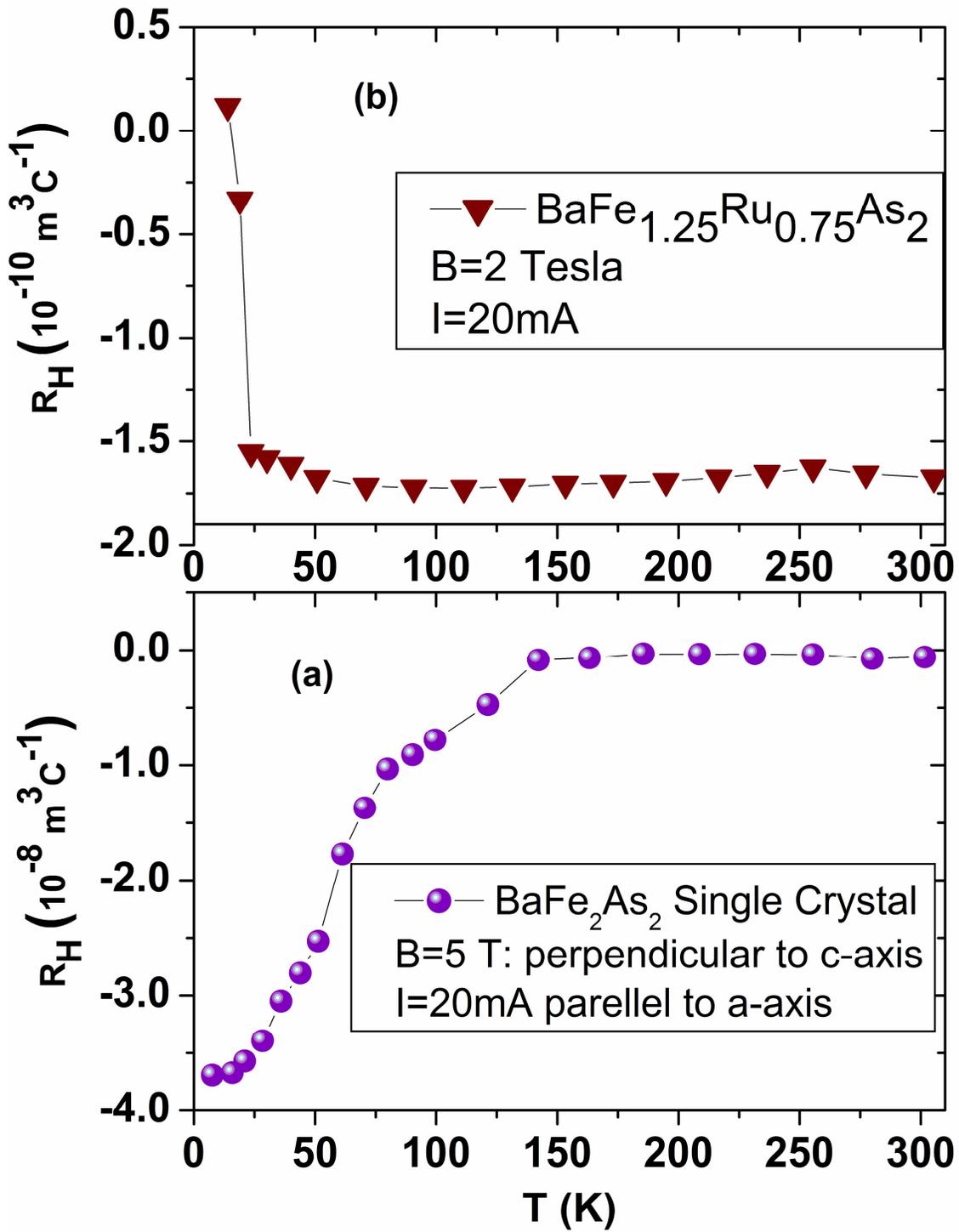

Fig.7

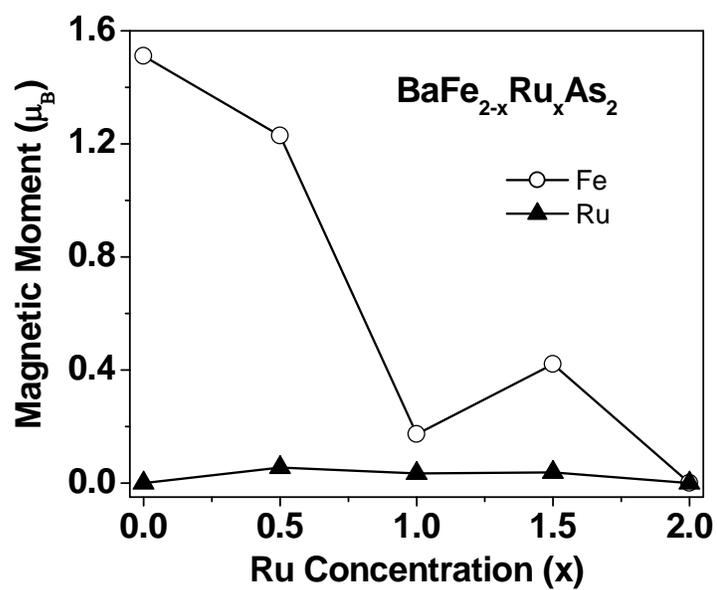

Fig.8

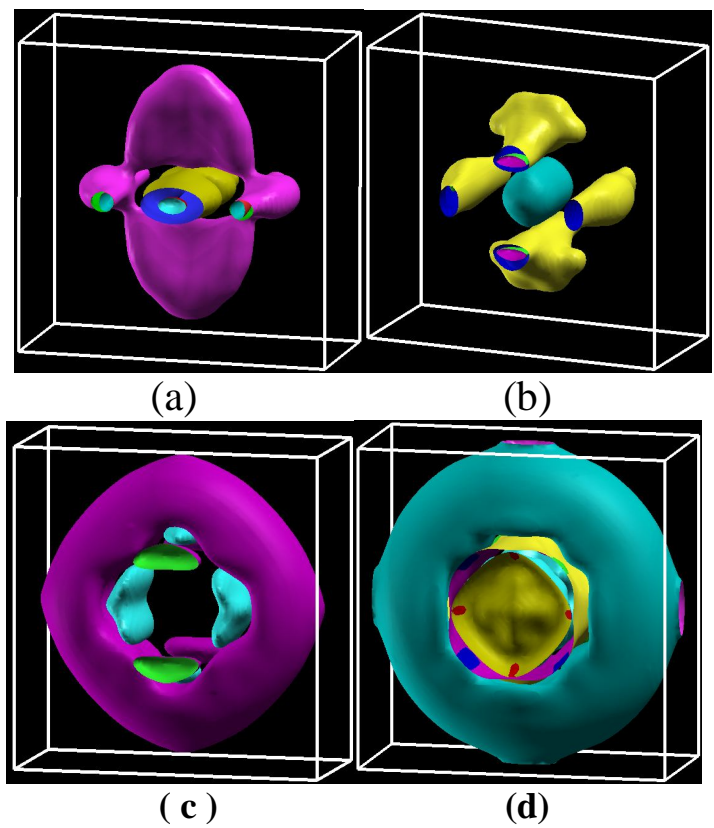

Fig.9

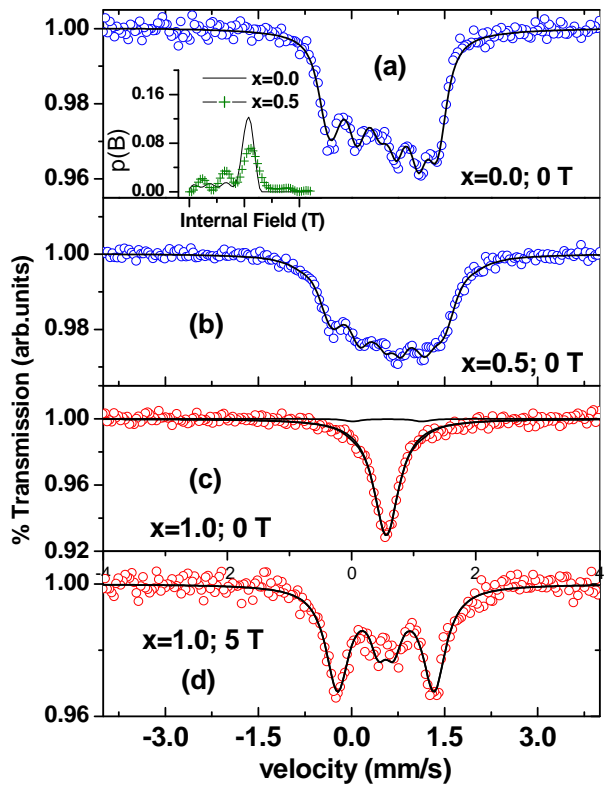

Fig.10